\documentstyle[art10]{article}
\begin{document}
\title{CO molecules in the shells of quasars 1556+3517 and 0840+3633 }
\author{Dubrovich V.K., Lipovka A.A.}
\maketitle

\begin{abstract}
New class of low- ionization BAL quasars (Becker et al. 1997),
represented by QSO 1556+3517 , 0840+3633 and 0059-2735, is considered.
Their optical spectra are completely dominated by absorption
features, whereas emission lines are weak or absent. Possibility
to explain these exclusive features of their spectra by existence
of molecular shells, responsible for the absorption is discussed.
In particular, it is suggested to identify line at $\lambda$ = 2063 A,
attributed to CrII, as a line of Cameron band
$a^{3}\Pi_{i} \leftarrow X^{1}\Sigma^{+}$ of CO molecule.
Estimation of the expected optical depth value in rotational radio
lines of this molecule is made. It is suggested to carry out
radio observations of quasars 1556+3517 and 0840+3633 to check the
molecular hypotheses.
\end{abstract}

\section{Introduction}

Investigation of absorption features in spectra of remote
quasars is the most effective method to explore the physical conditions
in absorbers. In this way one can obtain data not only on the galaxies
at line of sight quasar- observer, but also on shell of the quasar.

Last time the high activity take place in the field of searching and
identification of molecular lines in spectra of quasars.
Possible existence of this lines, created in distant molecular clouds,
was first suggested by Carlson (1974) for quasar 4C05.34 , in spectrum
of which he was trying to identify lines of $H_{2}$ molecule. However
these lines was certainly detected by Levshakov and Varshalovich (1985)
in absorption spectrum of the quasar PKS 0528-250 at Z = 2.811. This
identification was confirmed later by higher-resolution observations
independently by Foltz, Chaffee and Black (1988) and by Cowie and
Songaila (1995).

First possible detection of CO UV- lines in a quasar spectrum was made by
Varshalovich and Levshakov (1979) in spectrum of source PHL957
($z_{c}$ = 2.69), but identification of these lines was not confirmed
up to now. Only the upper limits were estimated by
Levshakov, Folts, Chaffee, Black (1989),
Levshakov, Folts, Chaffee, Black (1992), and
Lu, Sargent, Barlow (1997).

Possible radio observations of molecular clouds at cosmological distances
was discussed by Khersonsky, Varshalovich, Levshakov (1981).
From year to year, more and more distant radio- absorption system of CO
molecules in spectra of quasars are observed. The most remote CO-
absorber, recently discovered, is located at Z = 4.6.

The observations of the
radio lines of CO molecule in galaxies located at line of sight,
became usual events, whereas existence of this molecule in the
quasar shell is not believed to be probable due to strong ionized
radiation of the quasar core. However recently Guilloteau et al (1997)
observed rotational transition $J'J$ = 5-4 of the CO molecule at
$z_{a}$ = 4.407 in spectrum of radioquiet quasar BRI 1335-0415.
This result is of particular interest because, as it was mentioned by
the authors, this line can originate in the gas of the QSO host galaxy
(red shift of the quasar, determined with the number of lines
of atoms and ions is estimated to be from 4.407 to 4.41).

Recently we have discussed possible existence of molecular ion
$HeH^{+}$ in the hot quasar shell (Dubrovich and Lipovka 1995).
To form $HeH^{+}$, the ionized hydrogen (and hence strong ionized
radiation) is required, but in spite of rather small ionization potential
of this molecule, even under this circumstances the optical depth
in rotational lines of this molecule can be sufficiently large to be
detected. In relation to this topic the new class of iron- lowionized
BAL quasars, suggested by Becker et al. (1997) is of particular
interest, because in their shells the molecules like CO could possibly
survive.

This population, referred to as "iron Lo BAL QSO", is represented by
three objects:
1556+3517, 0840+3633 and 0059-2735 with redshifts z= 1.48 , 1.22 and 1.62
respectively. Spectra of these quasars are heavily attenuated at optical
wavelength in region less than 2800 A in the quasar restframe.
The number of ions and nutrals were found in their spectra (in particular
$FeI$ and $MnI$ characterized by ionization potentials 7.9 and 7.4 eV).
Large optical depths of nutral $Fe$ and $Mn$ in spectrum of 1556+3517
on the one hand, and unusually strong absorption in lines and continuum
as well, on the other hand, lead to conclusion on low degree of
ionization in their shells. It mean that strong ionized radiation of
their cores is shielded by dense internal part of the shell. It suggest
on the possible existence of steady molecules with high
(about 10 eV) dissociation energies in the shells.

Among suggested by Becker et al. (1997) identification, there is rather
strong line at
$\lambda \approx$ 2063 A, attributed to $CrII$. This identification is
not apparent due to uncertainty on the $Cr$ abundance in the shell
(which need not to be Solar one). In addition, it should be noted the main
depth in the $CrII$ line must grow in thin internal region because
of low ionization degree in the shell.

At the same time the transition from the ground state
$X^{1}\Sigma^{+}$ of CO molecule to the first exited one $a^{3}\Pi_{i}$,
has wavelength $ \lambda = 2062.9$ A, and next transition
$a^{'3}\Sigma \leftarrow X^{1}\Sigma^{+}$
has $ \lambda = 1806.5$ A (the line at this wavelength in the spectra of
the quasars is attributed to $SiII$ : $\lambda = 1808$ A).
CO molecule has dissociation energy $E_D = 11 eV$ ( it is more than value
of ionization potential of many atoms), and for this reason the possible
existence of CO in the quasars shells, discussed in this paper,
become very attractive. In this case one can identify the line at 2063 A,
as a line of Cameron band $a^{3}\Pi_{i} \leftarrow X^{1}\Sigma^{+}$ of
CO molecule.

If this molecular hypothesis is supported by observations of CO molecule
in the QSO shell, it will become a corner-stone in the model of
atmosphere of this objects. Besides that, it will be necessary to
reidentify the number of absorption lines in their spectra.

\section{Optical spectrum}

Two most prominent members of this population are quasars 1556 +3517
and 0840 +3633, distinguished by their strong absorption. Both
objects are redder than typical quasars at these redshifts.
So, their magnitudes taken from APM POSS I are O=21.2 , E=18.7
for 1556 +3517 and O=17.3, E=15.9 for 0840 +3633, whereas typical
values for intermediate- redshift quasars are O-E $\approx$ 0.5
(Becker et al. 1997). As a rule a reddening is connected with
existence of dust, but the spectra character does not support this
assumption.

There are three e-vibronic bands of CO molecule in the region of
wavelengths from $\lambda$ = 1400 to $\lambda$ = 3500 A
in which the spectra of quasars are shown in the paper of Backer
et al (1997). The most strong absorption band, should be mentioned first,
is $A^{1}\Pi \leftarrow X^{1}\Sigma^{+}$. It is allowed transition
characterized by oscillator strengths $\approx 10^{-2}$. It cover
the wavelength region shorter than 1544 A. Spectra of quasars
1556+3517 and 0840+3633 are totally absorbed in this region, and this fact
is an argument to support the molecular hypothesis.

The next term of our interest is lower triplet state $a^{3}\Pi_{i}$
of CO molecule. Cameron band $a^{3}\Pi_{i} \leftarrow X^{1}\Sigma^{+}$
is forbidden by the spin selection rule, however this transition can be
observed due to strong spin- orbital interaction of $a^{3}\Pi_{i}$ and
$A^{1}\Pi$ states (James 1971).
The first observation of this band was made by Barth et al. (1969) in
emission from upper atmosphere of Mars, but the oscillator strengths
(need for correct identification) was obtained latter by James (1971).
In consequence to these data, the values $f(v',v)$ for transitions
$a^{3}_{i}(v') \leftarrow X^{1}\Sigma^{+}(v)$ are
$f(1,0) = 2$ x $10^{-7}$ ( $\lambda_{10} = 1993 A$), and
$f(0,0) \approx f(0,1) = 1.6$ x $10^{-7}$ ( $\lambda_{00} = 2064 A,
\lambda_{01} = 2159 A$). It should be stressed, in spectra of quasars
under consideration, there are absorption lines at wavelengths
$\lambda_{10}$ , $\lambda_{01}$ and $\lambda_{00}$.
This fact justify our suppose on molecular nature of these lines.
Recently Minaev, Plachkevytch and Agren (1995) recalculated the oscillator
strengths, but their values does not change significantly. For transition
of our interest $(v',v) = (0,0)$ they suggest the value $f(0,0) =
1.4$ x $10^{-7}$, which will be used farther.

Third transition fall within the region under consideration
($a^{'3}\Sigma^{+}(v) \leftarrow X^{1}\Sigma^{+}(v)$) is forbidden too.
To our knowledge, its oscillator strengths was not calculated to
the moment, because of strong perturbations of $a^{'3}\Sigma^{+}$ term
by other ones. In this case the correct estimation of equivalent
width of lines become impossible.

\section{Optical depth and expected flux in radio continuum}

The value of optical depth in line $\lambda = 2063$ A is known.
For this reason we can exclude the column density of CO molecules
from consideration, and calculate the ratio of the radio line depth
to the optical one: $\tau_r / \tau_o$ = $\alpha_r / \alpha_o$, where
\begin{equation}\label{r1}
\alpha_\nu = \frac{A_{mn}c^2}{8\pi\nu_{mn}^{2}}\frac{g_m}{g_n}\frac{N_n}{\Delta\nu_D}\lbrace 1-e^{-\frac{h\nu_{mn}}{k T_{mn}}} \rbrace \ ,
\end{equation}
and for optical wavelengths:
\begin{equation}\label{r2}
\alpha_o = \frac{\pi e^2}{mc} \frac{N_0}{\Delta \nu_D} f_{00} \ .
\end{equation}
In this expressions $N_n$ - is the molecule density at level $n$,
$A_{mn}$ - is the Einstein coefficients, $N_0$ - is the molecule density
at the zeros vibrational level, $f_{00}$ - is the oscillator strength
for the transition $a^{3}_{i} (v'=0) \leftarrow X^{1}\Sigma^{+} (v=0)$,
$\Delta \nu_D$ - is the Doppler width of line.
By substituting these expressions, we obtain:
\begin{equation}\label{r3}
\frac{\tau_r}{\tau_o} = \frac{mc^3\nu_{opt}}{8\pi^2e^2\nu_{J'J}^{3}}\frac{g_{J'}}{g_J}\frac{n_J A_{J'J}}{n_0 f_{00}}\lbrace 1-e^{-\frac{h\nu_{J'J}}{k T_{J'J}}} \rbrace \ .
\end{equation}
Under reasonable assumption $n_0 \approx 1$ and $T_{J'J} = T_{CMBR}$,
one can obtain $\tau_r / \tau_o \approx 1.1$ x $10^3 n_J$ for lower
rotational levels, listed in table 1.
In the table their wavelengths recalculated for quasar 1556+3517 are given.

\begin{table}[htbp]
\caption{Rotational transitions of CO molecule in the ground state.}
\label{tspsig}
\begin{center}
\renewcommand{\arraystretch}{1.25}
\begin{tabular}{lll@{\hspace{0.5cm}}lll}
\hline
$J'J$  &  $\nu_0 [GHz]$  &  $A_{mn} [s^{-1}]$   &  $\lambda (1+z)  [mm]$ \\
\hline
0 1 & 115.271204 & 7.166(-8) & 6.4498259 \\
1 2 & 230.537974 & 7.874(-7) & 3.2249749 \\
2 3 & 345.79598  & 2.483(-6) & 2.1500511 \\
3 4 & 461.04081  & 6.094(-6) & 1.6126104 \\
4 5 & 576.26793  & 1.215(-5) & 1.2901624 \\
5 6 & 691.47298  & 2.126(-5) & 1.0752108 \\
\hline
\hline
\end{tabular}
\renewcommand{\arraystretch}{1}
\end{center}
\end{table}

The fluxes from quasar 1556+3517 at frequencies 1.4 and 5 GHz are 30.6
and 27.0 mJy respectively (Becker, et al. 1997). Spectral index in this
case is $\alpha$ = -0.1, that lead to expected flux $F$ = 20 mJy at the
frequency of line $(J'J) = (21)$ shifted at $(1+z)$: $\nu$ = 92.9 GHz.
For quasar 0840+3633, the flux at 1.4 GHz is 1.3 mJy. If we take
$\alpha$ = -0.1, we will obtain the expected flux $F$ = 0.8 mJy at
frequency of observation $\nu$ = 103.8 GHz.
Such fluxes can be observed with modern radiotelescopes.
For observations with IRAM 3mm receiver, for example, the time of
observation should be approximately 1 minute for QSO 1556+3517 and
10 hours for 0840+3633 (note that $\tau_r > 1$).

\section{Conclusion}

In the paper, the assumption on the presence of CO molecules in
lowionized shells of quasars of new population was considered.
The possibility is connected with the fact that on the one hand,
there are very large optical depth of the shell in the region of allowed
band $A^{1}\Pi \leftarrow X^{1}\Sigma^{+}$, on the other hand,
in their spectra there are number of absorption lines can be attributed
to band $a^{3}\Pi_{i} \leftarrow X^{1}\Sigma^{+}$.

The values of calculated fluxes in radio continuum from QSO 1556+3517,
and optical depth in radiolines CO molecule, are sufficiently large
to detect predicted lines.

\section{Acknowledgements}

The authors express their gratitude to Professor D.A. Varshalovich and Dr.
S.A. Levshakov for valuable remarks and helpful discussions.\\

\section{References}

Barth C.A. et al. //Science,(1969) v.165, p.1004\\
Becker R.H., et al. //Ap.J. (1997) v.479, p. L93\\
Carlson R.W. //Ap.J. (1974) v.190, N2, p.L99.\\
Cowie L.L. and Songaila A.A. //Ap.J. (1995), V.453, p.596\\
Dubrovich V.K., Lipovka A.A. //Astron. Astrophys. (1995) v.296, p.307\\
Foltz C.B., Chaffee F.H. and Black J.H. //Ap.J. (1988), V.324, p.267\\
Guilloteau S., Omont A., McMahon R.G., Cox P., Petitjean P.// Astron. Astrophys. (1997), v.328, p.L1\\
James T.S. //J. Chem. Phys.(1971) v.55, p.4118\\
Khersonsky V.K., Varshalovich D.A. Levshakov S.A. //Astron. Zh, (1981), V. 58, p. 29\\
Levshakov S.A. and Varshalovich D.A. //MNRAS, (1985), V.212, p.517\\
Levshakov S.A., Folts C.B., Chaffee F.H., Black J.H. //A.J. (1989), V.98, p.205\\
Levshakov S.A., Folts C.B., Chaffee F.H., Black J.H. //Astron. Astrophys. (1992), V.262, p.385\\
Lu L., Sargent W.L.W., Barlow T.A. // astro-ph/9711298 (1997)\\
Minaev B., Plachkevytch o., Agren H.//J. Chem. Soc. Faraday Trans. (1995) v.91, p.1729\\
Varshalovich D.A. Levshakov S.A. //Pisma Astron. Zh, (1979), V.5, 371.\\

\end{document}